\input{aipcheck}

\documentclass[
    ,final            
  ]
  {aipproc}

\layoutstyle{6x9}

\usepackage{graphicx,psfrag}
\usepackage{epsfig}
\usepackage{amsmath}
 \usepackage{amscd}
 \usepackage{amssymb}
\usepackage{epsf,float}

\def\beq{\begin{equation}}

\def\eps{{\epsilon}}

\begin{document}

\title{Deterministic Quantization by Dynamical Boundary Conditions}
\classification{03.65.-w;	       
11.15.Kc;	       
  11.25.Mj.	        
  }
\keywords      {Quantization; Time; Determinism; Compact Dimensions; AdS/CFT; Extra Dimension.}

\author{Donatello Dolce}{
  address={Johannes-Gutenberg Universit\"at,  D-55099  Mainz, Germany}
}

 \begin{abstract} 
	 We propose an unexplored quantization method. It is based on the assumption of dynamical space-time intrinsic periodicities for relativistic fields, which in turn can be regarded as dual to extra-dimensional fields. As a consequence we obtain a unified and consistent interpretation of Special Relativity and Quantum Mechanics in terms of Deterministic Geometrodynamics   \cite{Dolce:2009ce}. 
	\end{abstract}
 
 \maketitle

Under the de Broglie assumption, a free bosonic fields is characterized by a frequency $\bar \nu$ and the energy of the related quanta $ E = h \bar \nu$ depends on the inverse of the periods $T_t = 1 / \bar \nu $. 
 We now want  to fix the solutions of the relativistic wave equations by assuming Periodic Boundary Conditions (PBCs) $\Phi(t, \mathbf x) \equiv \Phi(t + T_t, \mathbf x)$ as constraint, in place of the usual fixed values of the fields at the time boundaries. 
 As investigated in detail in \cite{Dolce:2009ce}, the resulting non-trivial field theory turns out to be consistent with Special Relativity (SR) and  shows remarkable overlaps with  Quantum Mechanics (QM) under  many of its different formulations and for several nontrivial phenomena, see also  \cite{Dolce:2010zz}.  
Both choices of BCs minimize the relativistic action on the time boundaries, therefore they have the same formal validity. 
Naively, by  imposing time periodicity $T_t$ we get a discretization of the frequency spectrum, that is, according to de Broglie relation,  a quantization of the energy such that $E_n = n \bar E \equiv n h /T_t$.  
Because of the underlying Minkowski metric, this time periodicity induces a periodicity on modulo of the spatial dimensions  $ \lambda_x$ and, for massive fields where there exists a rest reference frame, on the proper time $T_\tau$. 
It turns out that \cite{Dolce:2009ce} the relation between these space-time periods is $ 1/T^{2}_\tau =  1/T^{2}_t - c^2 /  \lambda^{2}_x$. 
Similarly to the energy, we can define the momentum and the mass of the quanta of the field as $c \mathbf  {\bar p} \equiv h \hat{\mathbf n}/\lambda_x$ and ${\bar M} c^2 \equiv h / T_\tau$ respectively. 
Thus the relativistic dispersion relation $\bar E^2 (\mathbf {\bar p}) = \bar M^2 c^2 + \mathbf {\bar p}^2 c^4$ is satisfied. 
 Hence, the momentum spectrum is $\mathbf p_n = n \mathbf {\bar p}  $. 
In this way we see that the  mass $\bar M$  of a field is fixed by the proper time periodicity $T_\tau$ which in turn represents the upper limit of the time periodicity  $T_t (\mathbf {\bar{p}})$.
 Indeed the theory describes the same energy spectrum as the ordinary normal ordered second quantization. 
 Therefore they are consistent with Lorentz transformations and we can introduce the covariant notation $c \bar p_\mu = h / T^\mu$ where $c T^\mu = (c T_t,  \lambda_x \hat{\mathbf n} )$. 
 The resulting space-time periodicities $T_\mu$ are  those of the ordinary de Broglie waves. 
  In analogy with the Matsubara theory, the periodicities $T_\mu$ must be regarded as dynamical since they are related to  the fundamental four-momentum $\bar p_\mu$. 
  On the contrary, in the Kaluza-Klein (KK) theory the  periodicity along the eXtra-Dimension (XD) is static because it is related to an invariant mass. 
 The worldline parameter $s = c \tau$ has invariant periodicity $\lambda_s = c T_\tau$ and can be thought of as being a \textit{virtual} XD  of a five dimensional  massless field: $d S^2 = c^2 d t^2 - d \mathbf x^2 - c^2 d \tau^2 \equiv 0$. 
 In fact the proper time periodicity $T_\tau$ yields the quantization of the energy spectrum in the rest frame, i.e. the KK mass tower $E_n(0)/c^{2} = n {\bar M}$. 
 Indeed, this field theory can be thought of as being dual to the KK field theory \cite{Dolce:2009ce}. 
  The worldline periodicity $\lambda_s$  is nothing else than the Compton wavelength of the field; it corresponds to the periodicity along the \textit{virtual} XD.  
 Since in the KK theory there are no tachyons, all the energy eigenmodes have positive defined energies. 
  Considering  the actual experimental time resolution \cite{Broglie:1925}, even for a bosonic field at rest with the mass of an electron this intrinsic time periodicity is already extremely fast,  $T_t \lesssim 10^{-20} s$. 
  On the other hand, massless fields  have infinite time periodicity in the limit of low fundamental momentum $\mathbf {\bar p}$.

 The concept of time arising from this model  is consistent with SR, essentially because PBCs minimize the relativistic action. 
We  can as usual solve  the Green function and note that, by turning on a source in the origin, it induces a retarded variation of the energy in any given point. 
Thus, just by energy conservation, a field in the interaction point passes  dynamically from a periodic regime to another periodic regime, allowing time ordering.

	We now study the mechanics of these periodic fields, showing the analogies with ordinary QM. For simplicity we assume only one spatial dimension. 
	The first thing we note is that these  fields are a sum of on-shell standing waves, that is
	 $
  \Phi( x,t) =   \sum_{n} A_n  \phi_n( x) u_{n}(t)  = \sum_{n}  A_n  \exp[{-i (E_n t - p_n  x)/\hbar}] 
  $. 
  The  Fourier coefficients $A_n$ ($n \in \mathbb{Z}$) describe the occupation distribution of the different energy levels.  
Actually, we are in  the typical case where a Hilbert space can be defined. 
	In fact,  the energy eigenmodes $\phi_n(x)$ form a complete set with respect to the  inner pro\-duct $\left\langle \phi | \chi \right\rangle \equiv \int_{0}^{ \lambda_{x}} {d x}  \phi^* ( x) \chi( x) / {\lambda_{x}}$. 	 
	 Furthermore, the time evolution is described by the equations of motion $(\partial_t^2 + \omega_n^2)u_{n}(t) = 0 $ which can be reduced to first order differential equations  $i \hbar \partial_t \phi_{n}({x}, t) = E_n \phi_{n}({x}, t)$.   
	 Formally, from the  Hilbert eigenstates $\left\langle {x}| \phi_n  \right\rangle_{} \equiv { \phi_n({x}) }/{\sqrt{\lambda_{x}}}$  we can build the Hamiltonian operator $\hat{H} \left| \phi_n \right\rangle_{}  \equiv E_n \left| \phi_n \right\rangle_{}$ and a momentum operator $\hat{p} \left| \phi_n \right\rangle_{}  \equiv p_n \left| \phi_n \right\rangle_{}$. 
{Defining generically $| \psi  \rangle \equiv \sum_n a_n |\phi_n\rangle$,  we have  nothing else than the Schr\"odinger equation $i \hbar \partial_t |\psi\rangle = \hat H |\psi\rangle$.} 
The time evolution operator $ U(t'; t) = \exp[{-{i}\hat{H}(t-t')}/{\hbar} ]$  is  Markovian: $U(t'';t') = \prod_{m=0}^{N-1} U(t'+  t_{m+1}; t' + t_{m} -  \epsilon)$, where  $N \eps = t'' - t'$. 
	Therefore, without any further assumption than  periodicity and following the standard procedure, we use the completeness relation in combination with the elementary Markovian time evolutions obtaining formally the usual Feynman path integral for a time independent Hamiltonian \cite{Dolce:2009ce}. 
 In fact,   the elementary space-time evolutions  are supposed to be on-shell also in the usual Feynman  formulation.
This fundamental result can be interpreted as a consequence of the fact that the invariance under space-time periods translation  allows a class of classical paths with different winding numbers that can interfere each other. 

A further fundamental overlap with ordinary QM  is given by the fact that, from the assumption of periodicity it is possible to extract the usual commutation relation $[x,\hat p] = i \hbar$  \cite{Dolce:2009ce} (\textit{e.g.} we may note that $[x, -i \hbar \partial_x]\Phi(x,t)= i \hbar \Phi(x,t)$). 
The  Heisenberg uncertainty relation is easily obtained  as a direct consequence of the periodic conditions  $E_n {R_{t}} = n\hbar$ which can be stated in a  Bohr-Sommerfeld form: in a given potential the allowed solutions are those with integer numbers of cycles. 
Following this recipe  it is easy to reproduce the usual solution of several  Schr\"odinger problems \cite{Dolce:2009ce}. 
Moreover,  generalizing  the symmetry breaking mechanism by BCs \cite{Csaki:2003dt} to a periodic electromagnetic  field at low temperature, we have an effective  quantization of the magnetic flux \cite{Dolce:2009ce} and other typical behaviors of  superconductivity \cite{Weinberg:1996kr}.  

 Indeed we have a  deterministic theory of relativistic waves where QM emerges due to periodic dynamics intrinsically so fast  that the system can only be  described statistically  \cite{Dolce:2009ce}. 
 At every observation it  turns out to be in a random phase of its apparently aleatoric evolution. 
	 This is just like observing  a dice rolling too fast to predict the result. 
	 In fact ``there is a close relationship between the quantum harmonic oscillator and the classical particle moving along a circle" \cite{'tHooft:2001ar}. 
	In our case there are \textit{not} local-hidden-variables as the time is a physical variable that can not be integrate out, and as the PBCs are an element of non locality. 
	 Therefore we can talk about \textit{determinism} since the present theory is not constrained by the Bell's or similar theorems.

	Because of the underlying dualism with a theory in XD, the interactions can be  intuitively formalized using a geometrodynamical approach that can be regarded as a generalization of AdS/CFT  \cite{Maldacena:1997re}. 
	 In fact, during interaction,  because of the variation of the 4-momenta  $\bar p^\mu$, the  space-time periodicities $T_\mu$ of the fields  are subject to deformations.  
	 This corresponds to  deforme  the flat space-time metric. 
	 For instance, the quark gluon plasma can be approximated as a volume of massless fields, with an exponential decay of the energy $\bar E \rightarrow e^{-k s} \bar E$ during the freeze-out \cite{Magas:2003yp}, \textit{i.e} an exponential and conformal dilatation of the space-time periodicities $T_\mu \rightarrow e^{k s}  T_\mu$.
	    The corresponding metric is indeed the (virtual) AdS one: $d x^\mu \rightarrow e^{-  k s} d x^\mu$.  
	 From the results obtained so far and the dualism with fields in XD, we expect that the classical dynamics of these periodic fields  in such a deformed background  reproduce the quantum behaviors  of the corresponding interaction scheme \cite{Dolce:2009ce}.  
	 Actually, as well know from AdS/QCD, classical dynamics in AdS reproduce with good approximation QCD behaviors \cite{Pomarol:2000hp}.

 Paraphrasing  Newton's law of inertia and the de Broglie hypothesis of periodicity, \textit{we assume that every isolated elementary system is described in terms of relativistic deterministic fields with dynamical space-time periodicities (as long as it doesn't interact)  $T_\mu = h / \bar p^\mu$}. 
 As much as Newton's law of inertia doesn't imply that every point particle moves in a straight line, our assumption of dynamical periodicity does not mean that the physical world should appear periodic.
 In fact there is not a single static periodicity which would serve as a privileged reference. 
 On the contrary elementary systems (fields) at different energies have different periodicities.
 
 Ordinary field theory is based upon de Broglie waves, which are then quantized by imposing commutation relations.  
To every de Broglie waves there is an associated  frequency $\bar v$ proportional to the energy $\bar E$, and thus an intrinsic periodicity $T_t(\mathbf  {\bar p}) = h / \bar E(\mathbf  {\bar p})$ usually called de Broglie \textit{internal clock}. 
It is important to note that time can  only be  defined by counting the number of cycles of  phenomena that is \textit{supposed} to be periodic, in order to ensure that the duration of a unit of time is always the same \cite{Einstein:1910}.   
Our usual - non compact - time axis is defined with reference to the Cs-133 atomic clock whose reference period is about $10^{-10} s$. An electron at rest has a de Broglie  \textit{internal clock} of about $\sim 10^{-20} s$ whereas an  hypothetical heavy particle of $1$ TeV mass has an \textit{internal clock} of $\sim 10^{-27} s$. 
A massless field such as the electromagnetic field (or the gravitational field) can in principle have  all  possible values of periodicities. 
In particular, depending on its energy, it can have an infinite period (or of the order of the age of the universe). 
Thus,  every value of our time axis is characterized by a \textit{unique} combination of phases of all the \textit{internal clocks} of the elementary fields constituting the system under investigation. 
This means that the external time axis can be  dropped. 
The flow of time can be effectively described using the ``ticks"  of the de Broglie \textit{internal clocks} (as in a calendar or in an stopwatch). 
Here, the massless fields provide the long time scales. 
This is a simplification:  these \textit{internal clocks} can vary their periodicities through interactions (exchange of energy),  their periods depend   on the reference system according to the relativistic laws, furthermore  the combination of two or more clocks with irrational ratio of periodicities results in ergodic (nearly chaotic) evolutions. 
Using  Einstein's words \cite{Einstein:1910}, to define a clock ``\textit{we must assume, by the principle of sufficient reason, that all that
happens in a given period is identical with all that happens in an arbitrary period}". 
Hence we can restrict our attention on the single periods all the de Broglie \textit{internal clocks} constituting our system. 
In other words this means that the physical information of the fields is contained in the single de Broglie periods $T_t$.  
Therefore, using the terminology of field theories in XD,  we formalize this assumption by saying that the fields can be  characterized by dynamical compact time intervals and  PBCs (or similarly Dirichlet BCs).
Similar arguments hold for the spatial dimensions.  
In the non relativistic limit, matter fields can be approximated to have infinite spatial periodicities and microscopic time compactifications proportional to its Compton wavelengths.  
Hence they can be regarded  as 3D objects. 
Furthermore, since they are spatially  localized inside their  microscopical Compton wavelengths, they can be effectively  regarded as non-relativistic 3D point-like particles. 
 Following these few logical steps we are lead to a formulation of relativistic fields in dynamical compact space-time dimensions with lengths $T_\mu$.

Another intuitive picture can be found in the many similarities with acoustic fields, \textit{i.e.} sets of standing waves generated by objects vibrating in compact spatial dimensions which determine their harmonic spectra (frequency eigenstates). 
  In a full relativistic generalization of the sound fields, the periodic fields can be thought of as being generated by vibrating objects (sources) characterized by intrinsically compact space-time dimensions. 
  Roughly speaking, massless fields at small momentum have a nearly infinite time periodicity (continuous spectrum) so that they behaves like fields propagating in an non-compact medium;  matter fields have microscopically compactified time dimension and they act like sources.  
  The difference to the usual field theory is that by imposing space-time periodicities $T_\mu$ as constraint we allow a ``timbre''  to the de Broglie waves, \textit{i.e.} a spectral composition. 
Remarkably, this assumption can potentially open an unexplored scenario where SR and  QM are unified in a \textit{deterministic} field theory.

\bibliographystyle{aipproc}

	 \end{document}